\documentclass[letterpaper,table]{zadpreprints}
\pdfoutput=1

\usepackage{ragged2e}
\sectionfont{\RaggedRight}

\newcommand{\eightpt}{\fontsize{7}{5}\selectfont}
\renewcommand{\footnotesize}{\eightpt}

\makeatletter 
\renewcommand{\@makefntext}[1]{%
  \setlength{\parindent}{0pt}%
  \begin{list}{}{\setlength{\labelwidth}{2mm}
    \setlength{\leftmargin}{40pt}%
    \setlength{\labelsep}{1pt}%
    \setlength{\itemsep}{0pt}%
    \setlength{\parsep}{0pt}%
    \setlength{\topsep}{0pt}%
    \color{black}\footnotesize}%
  \item[\@thefnmark\hfil]#1
  \end{list}%
}
\makeatother 

\graphicspath{ArXiV/}
\DeclareUnicodeCharacter{2212}{-}

\papertype{Commentary}

\title{Misplaced Confidence in Observed Power}

\citearticle{Rafi, Z. Misplaced Confidence in Observed Power. \textit{\href{https://arxiv.org/abs/1907.08242}{arXiv:1907.08242 [stat.ME]}} (2020).}

\author[1]{Zad Rafi \protect\orcidicon{0000-0003-1545-8199}}

\affil[1]{Department of Population Health, NYU Langone, New York, NY}

\corraddress{Zad Rafi \\ Department of Population Health, NYU Langone, New York, NY}
\corrcontact{$^{1}$Email: zad@lesslikely.com \\$^{1}$Twitter: \href{https://twitter.com/DailyZad}{@DailyZad} \\}

\hypersetup{
pdftitle={Misplaced Confidence in Observed Power },
pdfsubject={Applied Statistics},
pdfauthor={Zad Rafi},
pdfdate={2020-07-07},
pdfproducer=pdfLaTeX,
pdfcopyright={Copyright (C) 2020,  The Author(s)},
pdfkeywords={Statistical power, Confidence Intervals, Data Interpretation, Hypothesis Tests, \textit{P}-values, Statistical Significance, Evidence, Information, Statistical Methods, Research Methods, Methodology},
pdflang={en},
pdfmetalang={en}
}

\begin{document}
\maketitle

\begin{abstract}
\vspace{0.3cm}
\small
A recently published randomized controlled trial in \textit{JAMA} investigated the impact of the selective serotonin reuptake inhibitor, escitalopram, on the risk of major adverse events (MACE). The authors estimated a hazard ratio (HR) of 0.69 (95\% CI: 0.49, 0.96; $p$ = 0.03) and then attempted to calculate how much statistical power their study (test) had attained, and used this measure to assess how reliable their results were. Here, we discuss why this approach, along with other post-hoc power analyses, are highly misleading. 
\end{abstract}
\vspace{-.5cm}
\hspace{4cm}

\begin{multicols}{2}

\section{Background}
\vspace{-.2cm} 
\indent \hspace{0.5cm}
	Kim et al. \cite{kimEffectEscitalopramVs2018} present the data from a randomized trial showing that treatment with escitalopram for 24 weeks lowered the risk of major adverse cardiac events (MACE) in depressed patients following recent acute coronary syndrome. The authors should be commended for exploring new approaches to reduce major cardiac events. However, in their paper, they seem to misunderstand the purpose of statistical power. In statistical hypothesis testing, power is the pre-study probability of correctly rejecting the test hypothesis, such as a null hypothesis, given a correct alternative hypothesis \cite{greenlandStatisticalTestsValues2016}. This is a frequentist probability, so it is calculated with multiple replications of the same design in mind. Power analyses can be useful for designing studies but lose their utility after the study data have been obtained and analyzed \cite{Hoenig2001-vi}. \\ \indent
	In the design phase, Kim et al. \cite{kimEffectEscitalopramVs2018} used data from the Korea Acute Myocardial Infarction Registry (KAMIR) study \cite{leeBenefitEarlyStatin2011} and estimated the power of their design to be 70\% and 96\% for detecting 10\% and 15\% group differences, respectively, with a sample size of 300 participants. This is often referred to as “design power.” It is based on the study design and beliefs about the true population structure. After completing the study and analyzing the data, the authors attempted to calculate how much power their test had based on the estimated hazard ratio (HR = 0.69, 95\% CI: 0.49, 0.96; \textit{p} = 0.03). They estimated that their test had 89.7\% power to detect between-group differences in MACE incidence rates. This is often referred to as “observed power” because it is calculated from the estimated effect size from the study data \cite{Hoenig2001-vi}. This form of power analysis is \textit{extremely misleading}. \\ \indent
	One cannot know the true power of a statistical test because there is no way to be sure that the effect size estimate from the study (HR = 0.69) is the population effect size. Therefore, one cannot be certain that the \textit{observed power} (89.7\%) is the \textit{true power} of the test. Observed power can often mislead researchers into having a false sense of confidence in their results. Furthermore, observed power, which is a 1:1 function of the \textit{P}-value, yields no relevant information not already provided by the \textit{P}-value \cite{Hoenig2001-vi,Greenland2012-by}. This also remains true for studies that are pooled in meta-analysis and then analyzed for average statistical power, which has recently gained momentum. However, this yields unstable interval estimates in even the most ideal of conditions \cite{mcshaneAveragePowerCautionary2020}. \\ \indent
	Statistical power can be useful for designing studies, but once the data from a study have been produced, it is the effect sizes, the \textit{P}-values, and the confidence/compatibility intervals that contain insightful information. Observed power adds little to this and can be misleading \cite{Hoenig2001-vi,mcshaneAveragePowerCautionary2020,Greenland2012-by}. If one wishes to see how cautious they should be about their point estimate, then they should look at the width of the interval estimates and the parameter values they cover \cite{greenlandStatisticalTestsValues2016,rafiSemanticCognitiveTools2020}.
\vspace{-.25cm}
	
{\fontsize{6}{7.25}\selectfont{
\bibliographystyle{naturemag}
\bibliography{References.bib}}}	
\end{multicols}
\end{document}